\documentclass[prl,aps,floatfix,showpacs,twocolumn 
]{revtex4-1}
\usepackage{amssymb,graphicx}
\usepackage{dsfont}
\usepackage{rotating}
\usepackage{array}
\usepackage{float}
\usepackage{bm, rotating}
\usepackage{subfigure}
\usepackage{wasysym}
\usepackage{listings}

\def\t{^\mathrm{T}}
\def\s2{\sqrt{2}}
\def\sf2{\frac{\sqrt{2}}{2}}
 \def\p{^\prime}
 
\def\lk{\left(}
\def\rk{\right)}

\begin{document}

\title{ Local Gaussian operations can enhance continuous-variable entanglement
 distillation}
\author{ShengLi Zhang and Peter van Loock}\email{peter.vanloock@mpl.mpg.de}

\affiliation{Optical Quantum Information Theory Group, Max Planck Institute for the Science of Light, G\"unther-Scharowsky-Str.1/Bau 26, 91058 Erlangen, Germany}
\affiliation{Institute of Theoretical Physics I, Universit\"at Erlangen-N\"urnberg, Staudstr.7/B2, 91058 Erlangen, Germany}

\begin{abstract}
Entanglement distillation is a fundamental building block in long-distance quantum communication. Though known to be useless on their own for distilling Gaussian entangled states, local Gaussian operations may still help to improve non-Gaussian entanglement distillation schemes. Here
we show that by applying
local squeezing operations, both the performance and the efficiency of existing distillation protocols can be enhanced. We derive the optimal enhancement through local Gaussian unitaries, which can be obtained even in the most natural scenario when Gaussian mixed entangled states are shared after their distribution through a lossy-fiber communication channel.
\end{abstract}
\maketitle


Entangled quantum states can be
seen as one of the essential resources in quantum information processing (QIP)
\cite{horodecki09}. However, entanglement is fragile and easily degraded by
uncontrollable environment-induced noise, for instance, during its
distribution over longer distances.
In order to circumvent this problem, various entanglement distillation protocols have been proposed
capable of improving shared entanglement by local operations
and classical communication \cite{bennet}.
In the realm of continuous-variable QIP, although Gaussian
entangled states and Gaussian operations \cite{EisertPlenioReview,AdessoIlluminati}
are widely used for teleportation,
dense coding, and other protocols \cite{RMP}, distilling Gaussian entanglement
in the potentially most efficient way
through only Gaussian operations has been shown to be impossible \cite{nogo1,nogo2,nogo3}.

In the mean time, efforts have been made to incorporate the necessary non-Gaussian element
into optical entanglement distillation schemes. For example,  Opatrn\'{y} \emph{et al.}
introduced the photon subtraction (PS) strategy employing one of the most readily available
non-Gaussian operations to distill stronger entanglement \cite{Opatrny}. This type of distillation was recently demonstrated experimentally \cite{NatPhotonic}. In other, complementary experiments, non-Gaussian entangled states
were initially prepared and subsequently, since no longer restricted by the above no-go results,  distilled
through Gaussian operations \cite{Hage08,Dong}. All these results and developments seem to imply
that local Gaussian operations are useless for the distillation of Gaussian entangled states (see ,for example,
Ref.~\cite{fiur2010}; however, note \cite{footnote1}).

In the present work, we shall demonstrate that Gaussian operations, though insufficient for distilling
Gaussian entanglement on their own, can still be used
to enhance existing non-Gaussian distillation schemes. More specifically,
we present examples of a distillation protocol that makes use of
local squeezing operations in addition to Opatrn\'{y} \emph{et al.'s} non-Gaussian PS strategy.
We then show that after distillation, both the success probability and the entanglement (in terms of
logarithmic negativity) are improved in a regime when an initial Gaussian two-mode
squeezed vacuum state (TMSS) is weakly squeezed. Moreover, this result even holds when
the TMSS is first subject to an amplitude damping channel, as it would be the case,
for instance, during an optical-fiber-based entanglement distribution. In fact, we are able to show
that the local Gaussian pre-processing prior to PS-based distillation (but after the channel
transmission) may become even more important for a lossy channel.

Our result is of both conceptual and practical significance. From a more fundamental point of view, it shows that in order to optimize quantum information tasks as important as entanglement distillation, discrete-variable and continuous-variable techniques should go hand in hand \cite{footnote2}. Since the PS technique \cite{Zavatta} as well as the implementation of online squeezing \cite{furusawaDemonst} is becoming state of the art, our protocol is of practical relevance too.

\begin{figure}[b]
  \includegraphics[width=8 cm]{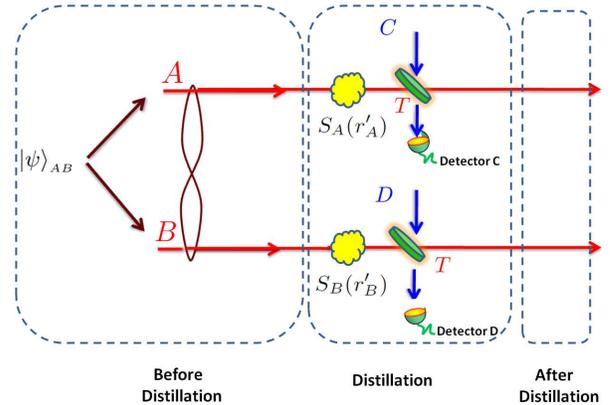}\\
  \caption{Distillation of Gaussian continuous-variable entanglement through PS and local squeezing,
  $S_A(r_A^\prime)$ and $S_B(r_B^\prime)$. Here,
 $|\psi\rangle_{AB}$ is a pure TMSS.
Beam splitters with transmission coefficient $T$
and on-off
detectors $C$ and $D$ are used to realize the corresponding non-Gaussian PS operations.
}\label{LocalDet}
\end{figure}

{\it Preliminaries--} for representing a continuous-variable system consisting of $N$ bosonic modes, we shall employ  the phase-space description, where every mode $k$ can be conveniently expressed by field quadrature position and momentum operators, $\hat{x}_k=(\hat{a}_k+\hat{a}_k^\dagger)/\sqrt{2}, \hat{p}_k=(\hat{a}_k-\hat{a}_k^\dagger)/(i\sqrt{2})$, with $\hat{a}_k , \hat{a}_k^\dagger$ being the mode annihilation and creators operators. Throughout, we use hats to denote operators in Hilbert space. The canonical commutation relations for an $N$-mode system can be conveniently written as $[\hat{X}_m,\hat{X}_n]=i\Omega_{mn}$, $1\le m,n \le 2N$, with $\hat{X}\equiv (\hat{x}_1, \hat{p}_1,\cdots,\hat{x}_N, \hat{p}_N)$ and $\Omega=\bigoplus _{k=1}^N
\left(
\begin{array}{cc}
0 & 1 \\
-1 & 0
\end{array}
\right).$
Furthermore, the density matrix of the above $N$-mode system $\rho$ in infinite-dimensional Hilbert space is represented by the characteristic function $\chi(\xi)
=\mathrm{Tr}\{\rho\exp[i\hat{X}\t\xi]\},\xi\in\mathbb{R}^{2N}$, or equivalently by its Fourier transform, i.e., the Wigner function $
W(\mathbf{x})=\int_{\mathbb{R}^{2N}}\frac{d^{2N}
\xi}{{(2\pi)}^{2N}}\exp\left[-i\mathbf{x}\t
\xi\right]\chi(\xi).
$
In particular, Gaussian states are those states whose characteristic or Wigner functions are Gaussian:
$
\chi (\xi) =\exp \left[ -\frac{1}{2}\xi \t%
V\xi+i\mathbf{\bar{x}}\t \xi\right]
$, and
$
W(\mathbf{x},\mathbf{\bar{x}},V)=\frac{\exp\left[-\frac{1}{2}(\mathbf{x}-\mathbf{\bar{x}})\t V^{-1}(\mathbf{x}-\mathbf{\bar{x}})\right]}{(2\pi)^N\sqrt{\mathrm{det}V}}
$.
Such Gaussian states are fully characterized by displacements $\mathbf{\bar{x}}=\mathrm{Tr}(\hat{X}\rho )$ and a covariance
matrix (CM) $V$, with entries $V_{lm}
=\frac{1}{2}\langle \hat{X}_l\hat{X}_m+\hat{X}_m\hat{X}_l\rangle-\langle \hat{X}_l\rangle\langle\hat{X}_m\rangle.
$ For example, for the pure TMSS, we have $\mathbf{\bar{x}}=0$ and the CM is given by
\begin{eqnarray}
V_{AB}=\frac{1}{2} \left(
\begin{array}
{cccc}
c&&s&\\
&c&&-s\\
s&&c&\\
&-s&&c
\end{array}\right),\label{Vab}
\end{eqnarray}
with $c=\cosh 2r, s=\sinh 2r$. Unitary state evolutions in Hilbert space generated
by Hamiltonians quadratic in the canonical operators correspond to symplectic
transformations, $S\in Sp(2N,\mathbb{R})$, in phase space.
In what follows, we will frequently use the symplectic single-mode squeezing,
$S(r)=\mathrm{diag}\{e^r,e^{-r}\}$, and two-mode beam-splitting operations.
The latter is performed via the $4\times 4$ matrix $S_{\mathrm{BS}}^{kl}$
with non-vanishing elements $(S_{\mathrm{BS}}^{kl})_{ii}= \sqrt{T}$, $\forall i$,
and $(S_{\mathrm{BS}}^{kl})_{31}=(S_{\mathrm{BS}}^{kl})_{42}=
-(S_{\mathrm{BS}}^{kl})_{13}=-(S_{\mathrm{BS}}^{kl})_{24}=\sqrt{1-T}$.
Here, $T$ represents the transmission coefficient and $kl$ the
two input modes of the beam splitter.

{\it Pure-state distillation--} with the notations above, we can now proceed to derive the state evolution in our modified entanglement distillation protocol, as shown in Fig.~\ref{LocalDet}.
As a start, let us first assume that the initial state is the pure TMSS,
$
|\psi\rangle_{AB}=\sum\limits_{n=0}^\infty \sqrt{1-\lambda^2}\lambda^n
|n\rangle_A |n\rangle_B,
\lambda=\tanh r,
$
with $|n\rangle$ denoting the Fock basis and $r$ the squeezing parameter.
To perform entanglement distillation, we propose to use two local squeezing operations $S_A(r_A^\prime)$ and $S_B(r_B^\prime)$ (with squeezing parameters $r_A^\prime$ and $r_B^\prime$) before the PS operation.
For the latter, we employ a beam splitter (with transmission coefficient $T$) and
conventional on-off detectors \cite{Oliver}. Throughout this paper, we refer to a successful distillation
event when both detectors register nonzero photon counts.

Now assume modes $C$ and $D$ are initially prepared in a vacuum state. Thus, the CM of
the four-mode state ABCD (before the photon detections) becomes
\begin{eqnarray}
V_{ABCD} &&=S_{\mathrm{tot}}
\left(V_{AB}\oplus \frac{1}{2}I_{CD}\right)
S_{\mathrm{tot}}\t,\\
S_{\mathrm{tot}} &&=\left[S_{\mathrm{BS}}^{AC}\oplus S_{\mathrm{BS}}^{BD}\right]
\left[S_A(r^\prime)\oplus S_B(r^\prime) \right],
\end{eqnarray}
under the additional assumption that optimal distillation will occur
for an initial symmetric state through symmetric local PS and squeezing operations,
$r_A^\prime=r_B^\prime=r^\prime$.

In our scheme, we propose to use on-off photon detectors, as commonly employed in quantum optics experiments.
Such a detector is represented by two measurement outcomes: `off' when no photons are detected and `on'
when one or more photons are detected. Through a successful distillation event, modes $C$ and $D$ are projected onto non-vacuum components and the state of modes $A$ and $B$ is reduced to
$
\widetilde\rho=\mathrm{Tr}_{_{CD}}\left[\rho_{_{ABCD}}I_{_{AB}}\otimes\hat{\Pi}_{_C}^{(on)}
\otimes\hat{\Pi}_{_D}^{(on)}\right]/P_{\mathrm{succ}}
$, with $\hat{\Pi}^{(on)}=I_\infty-|0\rangle\langle 0|=\sum_{n=1}^\infty |n \rangle\langle n|$. Throughout, we use $I_m$ to represent an $m$-dimensional identity matrix.
In order to obtain analytical results, we shall again employ the phase-space formalism. In fact, although
the single-mode operator $\hat{\Pi}^{(on)}$ leads to a non-Gaussian Wigner function, $W(\mathbf{x})=\frac{1}{2\pi}-\frac{1}{\pi}\exp[-\mathbf{x}^T I \mathbf{x}]$ \cite{PSwigner}, by expressing every single operator through a Wigner function and carrying out the corresponding integrals, we find that the Wigner function of the distilled state is a linear combination of four Gaussian functions:
 \begin{eqnarray}
&&W_{\widetilde{\rho}}(\mathbf{x})\cdot P_{\mathrm{succ}}=\sum_{j=1}^4 P_j W(\mathbf{x},\mathbf{0},V_j)\,,
\label{wignerOur}
\end{eqnarray}
with
$P_1=1$, $P_2=-[\mathrm{det }\left(V_C+I_2/2\right)]^{-1/2}$,
$P_3=-[\mathrm{det }\left(V_D+I_2/2\right)]^{-1/2}$,
$P_4=[\mathrm{det }\left(\Gamma_{CD}+I_4/2\right)]^{-1/2}$,
$V_1=\Gamma_{AB}$,
$V_2=\Gamma_{AB}-\sigma_1\left(V_C+I_2/2\right)^{-1}\sigma_1\t$,
$V_3=\Gamma_{AB}-\sigma_2\left(V_D+I_2/2\right)^{-1}\sigma_2\t$,
$V_4=\Gamma_{AB}-\sigma \left(\Gamma_{CD}+I_4/2\right)^{-1}\sigma\t$,
and $\Gamma_{AB},\Gamma_{CD},V_C,V_D$ defined by partitioning $V_{ABCD}$ as
\begin{eqnarray}
V_{ABCD}=
\left(\begin{array}
{cc}
\Gamma_{AB} & \sigma\\
\sigma\t & \Gamma_{CD}
\end{array}
\right)\,,\,\,\Gamma_{CD}=\left(
\begin{array}{cc}
V_C & \varsigma \\
\varsigma\t  & V_D
\end{array}
\right).\nonumber
\end{eqnarray}
Here, $
\sigma=\left(
\begin{array}{cc}
\sigma_1 ,  \sigma_2\\
\end{array}
\right)
$, and
$\sigma_1$ and $\sigma_2$ are both $4\times 2$ matrices, $\Gamma_{AB}, \Gamma_{CD}$ are $4\times 4$ matrices, and $V_C, V_D, \varsigma$ are $2\times 2$ matrices.
By also taking into account the normalization of the Wigner function and integrating both sides of Eq.~(\ref{wignerOur}) over the whole phase space,
the probability of successful distillation can be found to be given by $
P_{\mathrm{succ}}=\sum_{j=1}^4 P_j$.

Let us now use the logarithmic negativity \cite{LN,Plenio,eisert01} as a figure of merit to quantify the entanglement of the distilled state. The logarithmic negativity of a bipartite state $\rho_{_{AB}}$ is defined as
$E_N(\rho_{AB})=\log_2 ||\rho_{AB}^{\mathrm{T}_A}||$
with $\rho_{AB}^{\mathrm{T}_A}$ being the partially
transposed density operator.
In order to compute the logarithmic negativity, we shall
calculate the density matrix from notions in phase space. Using a method similar to Ref.~\cite{chen}, we have the following theorem \cite{errorchen}.

\noindent\textbf{Theorem 1:} let $V$ be the CM of a two-mode
zero-displacement Gaussian state and
$ \rho_{AB} $ be the corresponding
density matrix in the Fock basis, then the normalized matrix elements (for any $k_1,k_2,m_1,m_2=0,1,2,...$)
follow from
\begin{eqnarray}
&&\langle k_1 k_2 |\rho_{_{AB}} | m_1
m_2\rangle=
\frac{\partial_{t_1}^{k_1}\partial_{t_2}^{k_2}\partial_{t_1^\prime}^{m_1}\partial_{t_2^\prime}^{m_2}}{\sqrt{k_1! k_2!m_1! m_2! \det \Lambda}}\\
&&
\times\exp\left[\frac{1}{2}(t_1,t_2,t_1\p,t_2\p) M \lk\begin{array}{c}
t_1\\
t_2\\
t_1\p\\
t_2\p
\end{array}
\rk\right]\vline_{t_1=t_2=t_1\p=t_2\p=0}\nonumber,
\end{eqnarray}
 with
\begin{eqnarray}
 &&M=\sigma_x\otimes I_2+\sigma_z\otimes I_2 (L_2^* \Lambda^{-1} L_2^\dagger )\sigma_z\otimes I_2,\\
&& \Lambda= V+\frac{1}{2}L_2^{-1} (\sigma_x\otimes I_2)L_2^*,\\
&& L_2=\lk\begin{array}
{cccc}
-i\sf2& -\sf2 & & \\
&&-i\sf2 &-\sf2 \\
i\sf2 &-\sf2 & &\\
&&i\sf2 &-\sf2
\end{array}\rk,
\end{eqnarray}
where $\sigma_x,\sigma_z$ are the usual $2\times 2 $ Pauli matrices.

For our non-Gaussian state whose Wigner function is a linear combination of Gaussian functions, $W(\mathbf{x})=\sum_j P_j W(\mathbf{x},\mathbf{0},V_j)$, it can be easily proved that the density matrix follows the linear rule $\rho=\sum_j P_j \rho(V_j)$, with $\rho(V_j)$ determined by Theorem 1.

\begin{figure}
\centering
{
\includegraphics[width=8.5cm, height=6cm]{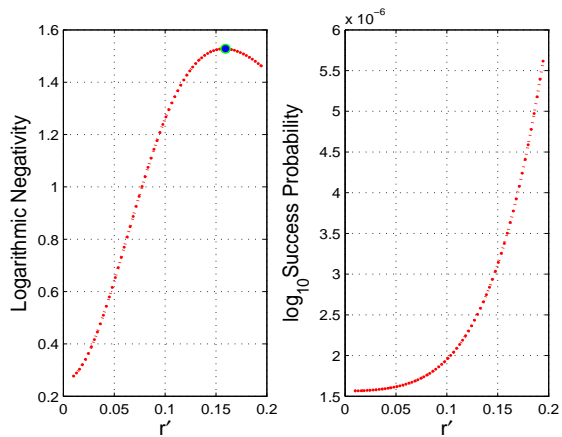}
}
\caption{
Distillation of a pure TMSS  ($r=0.025$) using local squeezers ($r^\prime$) and PS method ($T=0.95$). (a) logarithmic negativity of the output state; the blue circle indicates the optimal $r^\prime$ which maximizes the logarithmic negativity ($r^\prime_{\mathrm{opt}}=0.1565,E_N=1.5278$)
(b) success probability of distillation, i.e., the probability that both detectors obtain `on' results.
}
\label{FigPure}
\end{figure}

Using the above methods, we now present a numerical evaluation of the performance of our modified distillation protocol. Fig.~\ref{FigPure} shows the logarithmic negativity of the distilled state for $r=0.025,r^\prime \in [0.01, 0.20]$ \cite{Larger_r}. The logarithmic negativity attains a maximal value at an intermediate point $r^\prime_{\mathrm{opt}}=0.1565 $, whereas the success probability increases monotonically with local squeezing $r^\prime$. At the optimal point $r^\prime_{\mathrm{opt}}$, with success probability $3.5029\times 10^{-6}$ and $E_N=1.5278$, we obtain a significant improvement over Opatrn\'{y} 's original PS strategy (with $1.5645\times 10^{-6}$ and $E_N=0.1352$ \cite{Kitagawa,ourPaper}).
In other words, the local filter operations corresponding to the beam splitters and on-off detectors
(i.e., the local map for say mode $A$, $\mathcal{E}:\rho_A\rightarrow \sum_{i=1}^\infty \hat{E}_i\rho_A \hat{E}_i^\dagger/\mathrm{Tr}[\sum_i\hat{E}_i^\dagger\hat{E}_i
\rho_A]$, with $\hat{E}_i=\sum_{n=i}^\infty (-1)^i \sqrt{{n\choose i}}T^{(n-i)/2} (1-T)^{i/2} |n-i\rangle\langle n|$),
become more efficient when replaced by the set of operators $\{\hat{E}_i \hat{S}_{A}(r_A^\prime)\}_{i=1,2\cdots}$.
At the same time, local squeezing increases each mode's average photon
number and thus enhances the probability that photons are detected.

\begin{figure}[b]
\centering
{
\includegraphics[width=8.5cm, height=6cm]{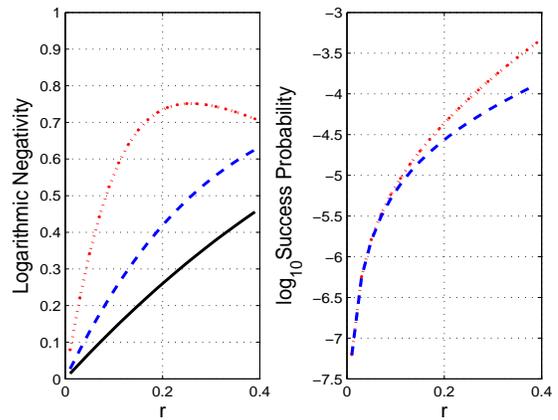}
}
\caption{
Mixed-state distillation with $\eta=0.5$. Comparing the distillation performances between Opatrn\'{y}'s PS strategy (blue, dashed line) and our local-squeezing enhanced PS strategy (red, dot dashed line), for $r \in [0.005, 0.4]$, $T=0.95$,
$r^\prime=r={\rm arctanh}(\lambda)$. The black (solid) line indicates the logarithmic negativity of the mixed state before distillation.
}
\label{FigMix0p8}
\end{figure}

{\it Mixed-state distillation--} in the above distillation scheme,
the local squeezers could be as well seen as part of the initial state preparation.
One could then argue that so far we have simply found the optimal Gaussian pure state
for PS-based entanglement concentration which turns out to be different from the TMSS.
It is therefore intriguing to examine whether the TMSS would also benefit
from local Gaussian pre-processing after its transmission through an imperfect
channel such as a lossy fiber. This would lead to a clear
distinction between local operations before and after the entanglement distribution,
corresponding to alternate state preparations or alternate state distillations, respectively.

For this purpose, we use beam splitters and auxiliary vacuum modes to model the optical amplitude-damping
channel \cite{ourPaper}. Our method introduced above still applies,
except for replacing
$
c\rightarrow c^\prime = 1-\eta+\eta\cosh 2r ,s\rightarrow s^\prime=\eta\sinh 2r
$
in Eq.~(\ref{Vab}), with $\eta$ being the transmission efficiency of the lossy channel. We find that local squeezing still helps to improve entanglement distillation of mixed states. In Fig.~\ref{FigMix0p8}, the logarithmic negativity and success probability
of distillation are shown. For simplicity, we consider the case when each mode of the TMSS is transmitted through a
$3 {\rm dB}$ ($\eta=0.5$) amplitude-damping channel. In the low-squeezing regime, e.g. for $r\in [0.005, 0.4]$, a significant improvement is obtained.

\begin{figure}[t]
\centering
{
\includegraphics[width=8.5cm, height=6cm]{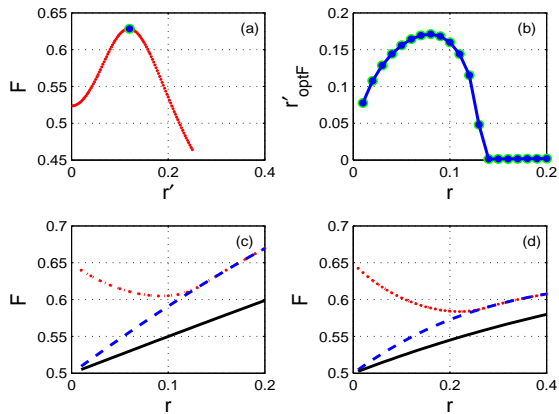}
}
\caption{ Teleportation fidelity. (a) with $\eta=1,r=0.025$ (b) optimal local squeezing $r_{\mathrm{optF}}^\prime$ as a function of $r$ (c) comparison between our modified scheme (red, dot dashed line) and Opatrn\'{y}'s PS strategy (blue, dashed line): the results coincide beyond $r\sim 0.13$ (d) using a distilled 3dB-amplitude-damped TMSS state; the threshold value becomes $r\sim 0.26$. The black (solid) line shows the fidelity of using the 3dB-damped TMSS state before distillation. Throughout we use $T=0.95$.
}
\label{teleFid}
\end{figure}

{\it Quantum teleportation--} the improvement in our modified scheme can be experimentally verified in an operational fashion through quantum teleportation \cite{Opatrny,Kitagawa}.
Let us consider standard unit-gain teleportation in which the entangled state after distillation is pre-shared and an unknown coherent state $|\alpha_0\rangle$ is to be teleported. The teleportation fidelity is $F_{\rm tel}= \langle\alpha_0|\rho_{\mathrm{out}}(\alpha_0)|\alpha_0\rangle$, with $\rho_{\mathrm{out}}(\alpha_0)$ being
the ensemble average of all output states conditioned upon different Bell
measurement results. Due to its linearity, the teleportation fidelity using the entangled state in Eq.~(\ref{wignerOur}) can be conveniently written as $F=\sum_{j=1}^4 P_jF_{\rm tel}(V_j)/P_{\mathrm{succ}}$, with (for any $\alpha_0$)
$F_{\rm tel}(V)=\det(R\alpha R +\gamma\t R+R\gamma+\beta+I_2 )^{-1/2}, R={\rm diag}(-1,1) $, where $\alpha,\beta,\gamma$
are defined by $V\equiv \lk\begin{array}{cc}
\alpha &\gamma\\
\gamma\t & \beta
\end{array}\rk$. In Fig.~\ref{teleFid}(a), we consider a pure TMSS state ($\eta=1,r=0.025$) being distilled and then used for teleporting an unknown coherent state. Fig.~\ref{teleFid}(b) shows the optimal local squeezing $r_{\mathrm{optF}}^\prime$ as a function of $r$. Clearly, $r_{\mathrm{optF}}^\prime$ begins to drop towards zero for a stronger pure TMSS state ($r\sim 0.13$), representing a threshold beyond which our modified scheme
ceases to improve Opatrn\'{y}'s PS strategy ($r^\prime=0$). In Fig.~\ref{teleFid}(c), the fidelity at $r^\prime=r_{\mathrm{optF}}^\prime$ is compared with Opatrn\'{y}'s PS strategy, while Fig.~\ref{teleFid}(d) shows the fidelity for using a distilled 3dB-damped TMSS state. In this case, a larger threshold value, $r\sim 0.26$, occurs. Thus, we find that for an amplitude-damped resource state, the local Gaussian pre-processing prior to distillation becomes even more significant.

{\it Optimal Gaussian unitaries--}
the most general symplectic transformation applicable to our local filters can be written as a sequence of phase rotation, local squeezing, and another phase rotation \cite{Braunstein},
$
U(r,\theta,\phi)= \left(\begin{array}{cc}
\cos \theta & \sin\theta\\
-\sin\theta & \cos \theta
\end{array}
\right) \left(\begin{array}{cc}
e^r & \\
  & e^{-r}
\end{array}
\right) \left(\begin{array}{cc}
\cos \phi & \sin\phi\\
-\sin \phi & \cos \phi
\end{array}
\right).$
For a fixed squeezing $r=\rm arctanh(\lambda) $ in the initial pure TMSS of Eq.~(\ref{Vab}), we applied two local unitary operations $U(r,\theta_A,\phi_A)$ and $U(r,\theta_B,\phi_B)$ to modes $A$ and $B$, respectively. We computed the logarithmic negativity and success probability for
3000 randomly chosen $\theta_A,\phi_A,\theta_B,\phi_B$, and we found that both figures of merit attain maximal values for $\theta_A=\phi_A=\theta_B=\phi_B=0$.

{\it Summary--} we demonstrated that local Gaussian operations can be useful to enhance
the distillation capabilities of non-Gaussian operations applied upon Gaussian entangled states,
even when the initial
states are Gaussian mixed states such as those emerging from an imperfect channel transmission for
realistic entanglement distribution in quantum communication. In our protocol, for the distribution of one
entangled-state copy, we considered a photon loss channel, and for one-copy distillation, we used the optimal
local unitary squeezers in addition to photon subtraction.
It remains an open question whether local Gaussian non-unitary maps
could further improve non-Gaussian entanglement distillation schemes.

SZ acknowledges support by the Max Planck Gesellschaft, Chinese Academy of
Sciences Joint Doctoral Promotion Programme (MPG-CAS-DPP), and Key Lab of Quantum
Information (CAS). PvL acknowledges support from the Emmy Noether Program of the DFG.


\begin{thebibliography}{99}
\bibitem{errorchen} note that there is an error in Eq.~(6) of Ref.~\cite{chen}, and it should correctly read $t_j t_j^\prime-t_j\mu_j+t_j^\prime \mu_j^*$ in that equation. Beyond this, we present a stronger theorem for which the density matrix obtained is automatically normalized.
\bibitem{Larger_r} in general, Theorem 1 applies to arbitrary CMs $V$, including arbitrary two-mode squeezing $r$. Here we choose initial squeezing $r=0.025$, although we could as well use larger squeezing $r$ and, with the same method, derive the optimal local squeezing $r_{\mathrm{opt}}^\prime$. However, for stronger squeezing $r$, $r_{\mathrm{opt}}^\prime$ will be so large that we have to keep many (e.g. $\ge 15$) photons in each mode to make sure that every CM $V_j(j=1,2,3,4)$ is transformed to a density matrix within a precision of $10^{-6}$. To double-check our computer program, we also transformed the density matrices back to the CMs and always kept enough photons in each mode to make sure that the elements of every newly obtained CM matrix $V_j^\prime$ are only negligibly different ($<10^{-6}$) from those of the original CM $V_j$. For example, the calculation of the density matrix and logarithmic negativity for $r=0.025, r^\prime=0.025$ takes only $12$ minutes and a $60$ MB memory, while only $10$ photons are kept in each mode, with our MATLAB program and Intel Core i5 $2.27$ GHz CPU. However, for larger r, say $r=0.04,r^\prime=0.10$, we have to consider up to $12$ photons corresponding to a computation time of $2.5$ hours with $100$ MB memory. Further, if we keep $14$ photons, the numbers become $15$ hours and $600$ MB memory, etc.
\end{thebibliography}
\end{document}